\documentclass{osa-article}

\journal{oe}
\articletype{Research Article}
\usepackage[T1]{fontenc}
\usepackage[latin9]{inputenc}
\usepackage{color}
\definecolor{note_fontcolor}{rgb}{0.800781, 0.800781, 0.800781}
\usepackage{verbatim}
\usepackage{textcomp}
\usepackage{bm}
\usepackage{amsmath}
\usepackage{graphicx}
\makeatletter

\usepackage[english]{babel}
\usepackage{dcolumn}
\usepackage{color}
\usepackage{subfigure}

\def\bra#1{\langle #1 |}
\def\ket#1{| #1 \rangle}

\def\A{\mathcal{A}}

\def\br{\bm{\rho}}

\makeatother

\begin{document}

\title{Correlation Plenoptic Imaging between Arbitrary Planes}

\author{Francesco Di Lena,\authormark{1}, Gianlorenzo Massaro,\authormark{2}, Alessandro Lupo,\authormark{2}, Augusto Garuccio,\authormark{2,1} Francesco V. Pepe,\authormark{2,1,*} and Milena D'Angelo\authormark{2,1,$\dagger$}}

\address{\authormark{1}INFN, Sezione di Bari, I-70125 Bari, Italy \\
\authormark{2}Dipartimento Interateneo di Fisica, Universit\`{a} degli Studi di Bari, I-70126 Bari, Italy }

\email{\authormark{*}francesco.pepe@ba.infn.it} \email{\authormark{$\dagger$}milena.dangelo@uniba.it} %% email address is required



\begin{abstract*}
We propose a novel method to perform plenoptic imaging at the diffraction limit by measuring second-order correlations of light between two reference planes, arbitrarily chosen, within the tridimensional scene of interest. We show that for both chaotic light and entangled-photon illumination, the protocol enables to change the focused planes, in post-processing, and to achieve an unprecedented combination of image resolution and depth of field. In particular, the depth of field results larger by a factor 3 with respect to previous correlation plenoptic imaging protocols, and by an order of magnitude with respect to standard imaging, while the resolution is kept at the diffraction limit. The results lead the way towards the development of compact designs for correlation plenoptic imaging devices based on chaotic light, as well as high-SNR plenoptic imaging devices based on entangled photon illumination, thus contributing to make correlation plenoptic imaging effectively competitive with commercial plenoptic devices.
\end{abstract*}

%%%%%%%%%%%%%%%%%%%%%%%%%%  body  %%%%%%%%%%%%%%%%%%%%%%%%%%

\section{Introduction}

Plenoptic imaging (PI) is a recently established optical imaging technique that allows to collect the \textit{light field}, namely, the composite information on spatial distribution and direction of light coming from the scene of interest \cite{adelson1992single,ng2005light}. The reconstruction of light paths can be used, in post-processing, to refocus out-of-focus planes, change the point of view and extend the depth of field (DOF) within the three-dimensional scene of interest. PI is also one of the simplest and fastest methods to obtain three-dimensional images with the current technology~\cite{broxton2013wave,xiao2013advances,prevedel2014simultaneous,
ren2017fast,dansereau2013decoding,adhikarla2015exploring,wanner2012globally}. In state-of-the-art plenoptic cameras, the composite information of spatial distribution and direction of light is collected by means of a microlens array; this imposes a significant resolution loss, well below the diffraction limit defined by the numerical aperture (NA) of the camera lens \cite{ng2005fourier,georgiev2012multifocus,goldlucke2015plenoptic}. 
Attempts to weaken the resolution vs. DOF trade-off have been made by using signal processing and deconvolution~\cite{broxton2013wave,prevedel2014simultaneous, waller2012phase,georgiev2006spatio,perez2014super}, and other algorithms and analysis tools \cite{dansereau2013decoding,li2016scalable}.

In this perspective, we have recently proposed a fundamentally different approach, named correlation plenoptic imaging (CPI), where spatio-temporal correlation properties of light are exploited to physically decouple the image formation from the retrieval of the propagation direction of light, which are registered by two disjoint sensors \cite{d2016correlation}.
As a consequence, no microlens array is required and diffraction-limited resolution can be recovered.
CPI has been proposed for both chaotic light~\cite{d2016correlation} and entangled photons illumination \cite{pepe2016correlation}, and several alternative configurations \cite{pepe2017exploring,scagliola2020correlation,dilena2018correlation} have been considered. The first experimental demonstration of CPI has been performed with chaotic light \cite{pepe2017diffraction}, and the analysis of the signal-to-noise ratio in specific cases \cite{scala2019signal,scala2020signal} have been performed.

The common feature of most plenoptic imaging protocols so far explored is the fact that directional information is retrieved by imaging two specific planes: one arbitrarily chosen within the 3D scene of interest, and one coinciding with either the focusing element or any other lens within the device.\footnote{This is not the case in so called Plenoptic 2.0, where the microlenses create redundant images of the scene of interest \cite{georgiev2009high} and propagation direction is obtained by a sort of triangulation, with the effect of improving the depth of field while further sacrificing the image resolution}. However, to focus composite lenses, such as camera lenses or microscope objectives, is not trivial and imposes the identification of correction factors to be introduced within the refocusing algorithm to account for the uncertainty about the effective distance between the two reference planes.

In this paper, we demonstrate that this difficulty can be overcome by performing plenoptic imaging starting from the acquisition of diffraction-limited images of two generic planes typically chosen within the tridimensional scene of interest. The core of this proposal, which we shall name \textit{correlation plenoptic imaging between arbitrary planes} (CPI-AP), is to employ correlated light, such as chaotic light or entangled photons, and to measure correlations between two disjoint sensors, placed in the conjugate planes of the two arbitrarily chosen planes. 
Besides highly simplifying the experimental implementation and improving the precision of refocusing, the proposed protocol has the further advantage of relieving the resolution versus DOF compromise, so as to  reach an unprecedented combination of these two parameters.
As we shall discuss, the area between the two chosen planes is also very interesting in terms of achievable resolution; it is thus intriguing to have the opportunity to choose the distance between the arbitrary planes based on both the extension of the sample and the required resolution of the overall 3D scene.

\section{CPI-AP with chaotic light illumination}
\label{sec:cnp}

\begin{figure}
\centering
\includegraphics[width=0.9\textwidth]{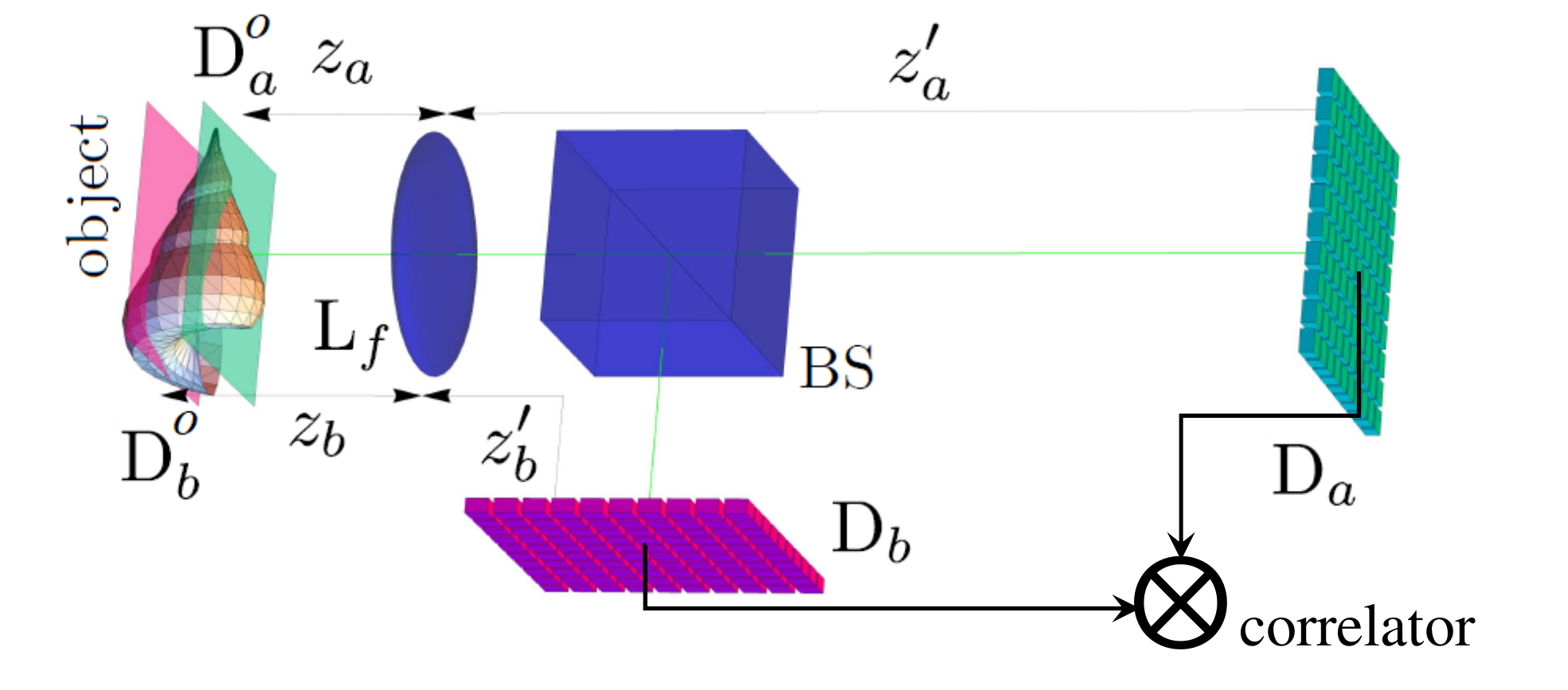}
\caption{Schematic representation of the CPI-AP protocol, in the case of choatic light illumination. The object is modeled as a chaotic light emitter. The lens $L_f$ generates the images of the two planes $\text{D}^o_a$ and $\text{D}^o_b$, arbitrarily chosen close to the three-dimensional object of interest, on the two spatially-resolving detectors $\text{D}_a$ and $\text{D}_b$, respectively. Combined information on the distribution and direction of light is retrieved by computing correlations of intensity fluctuations between each pair of pixels of the two detectors.}
\label{fig:cnp:setup}
\end{figure}

Let us start by analyzing the CPI-AP protocol in the case the illuminating light is emitted by a chaotic source. A schematic representation is reported in Fig.~\ref{fig:cnp:setup}. Light from the object passes through the lens $L_f$, of focal length $f$, and is separated by a beam splitter (BS) in two beams, each one detected by a different spatially-resolving sensor, $\mathrm{D}_a$ and $\mathrm{D}_b$. The detectors are placed in the conjugate planes of two planes arbitrarily chosen in the surrounding of the object, indicated with $\mathrm{D}^o_a$ and $\mathrm{D}^o_b$, respectively, and by the same color as their conjugate sensors; if $z_a'$ and $z_b'$ are the distances between the lens $L_f$ and the two sensors $\text{D}_a$ and $\text{D}_b$, respectively, the thin-lens equations 
\begin{equation}\label{eq:cnp:tl}
\frac{1}{z_j} + \frac{1}{z_j'} = \frac{1}{f}
\qquad \text{with } j= a,b
\end{equation}
define the distances $z_a$ and $z_b$ of the conjugate planes of the detectors, $\text{D}^o_a$ and $\text{D}^o_b$, from $L_f$. 
As we shall demonstrate, plenoptic information is contained in the spatio-temporal correlations characterizing the intensity fluctuations retrieved by the two sensors.
 
To simplify the computation, we shall consider a planar object, placed at a distance $z$ from $L_f$, whose emission is characterized by the light intensity profile $\A(\br_o)$ and by negligible transverse coherence. Though, for definiteness, the object will be treated as an emitter of chaotic light, the
working principle remains unchanged in case it either reflects, transmits or scatters chaotic light. As mentioned above, CPI-AP is based on the measurement of equal-time correlations between the light intensities measured at the points of planar coordinates $\br_a$, on detector $\mathrm{D}_a$, and $\br_b$, on detector $\mathrm{D}_b$. More specifically, the relevant information that enables plenoptic imaging is contained in the correlation between \textit{intensity fluctuations}
$
\Gamma(\bm{\rho}_a,\bm{\rho}_b) = \langle \Delta I_a (\br_a) \Delta I_b (\br_b) \rangle
= \langle I_a(\bm\rho_a) I_b(\bm\rho_b) \rangle -
\langle I_a(\bm\rho_a) \rangle \langle I_b(\bm\rho_b) \rangle $ . 
Under the assumption that the source is ergodic \cite{mandel1995optical}, the ensemble average appearing in $\Gamma(\bm{\rho}_a,\bm{\rho}_b)$ can be approximated by a time average, in line with the experimental procedure \cite{pepe2017diffraction}. 
In addition, we will consider quasi-monochromatic light of central wavelength $\lambda$ and wavenumber $k=2\pi/\lambda$, and propagate light from a generic object point $\br_o$ to a detector point $\br_a$ ($\br_b$) by means of the corresponding paraxial optical transfer functions \cite{goodman2005introduction}. We thus obtain the correlation function, which reads, up to irrelevant factors,
\begin{equation}\label{eq:cnp:gamma}
\Gamma(\bm\rho_a,\bm\rho_b)  = 
\left|\int d^2\bm\rho_o \A(\bm\rho_o) \int d^2\bm{\rho}_{\ell} P^*(\bm\rho_{\ell}) \int d^2\bm{\rho'}_{\ell} P(\bm\rho'_{\ell}) e^{-i k [ \phi_a(\bm\rho_o,\bm\rho_{\ell},\bm\rho_a) - \phi_b(\bm\rho_o,\bm\rho'_{\ell},\bm\rho_b) ]} \right|^2
\end{equation}
where
\begin{equation}\label{eq:cnp:phi}
\phi_j (\bm\rho_o,\bm\rho_{\ell},\bm\rho_j) =
\left(\frac{1}{z}-\frac{1}{z_j}\right)\frac{\bm\rho_{\ell}^2}{2}-
\left(\frac{\bm\rho_o}{z}-\frac{\bm\rho_j}{M_j z_j}\right) \cdot\bm\rho_{\ell} ,
\end{equation}
with $P(\br_{\ell})$ the lens pupil function and $M_j=-z_j'/z_j$ the magnifications of planes $\text{D}^o_j$ on sensors $\text{D}_j$, with $j=a,b$.

\begin{figure}
\centering
\includegraphics[width=0.9\textwidth]{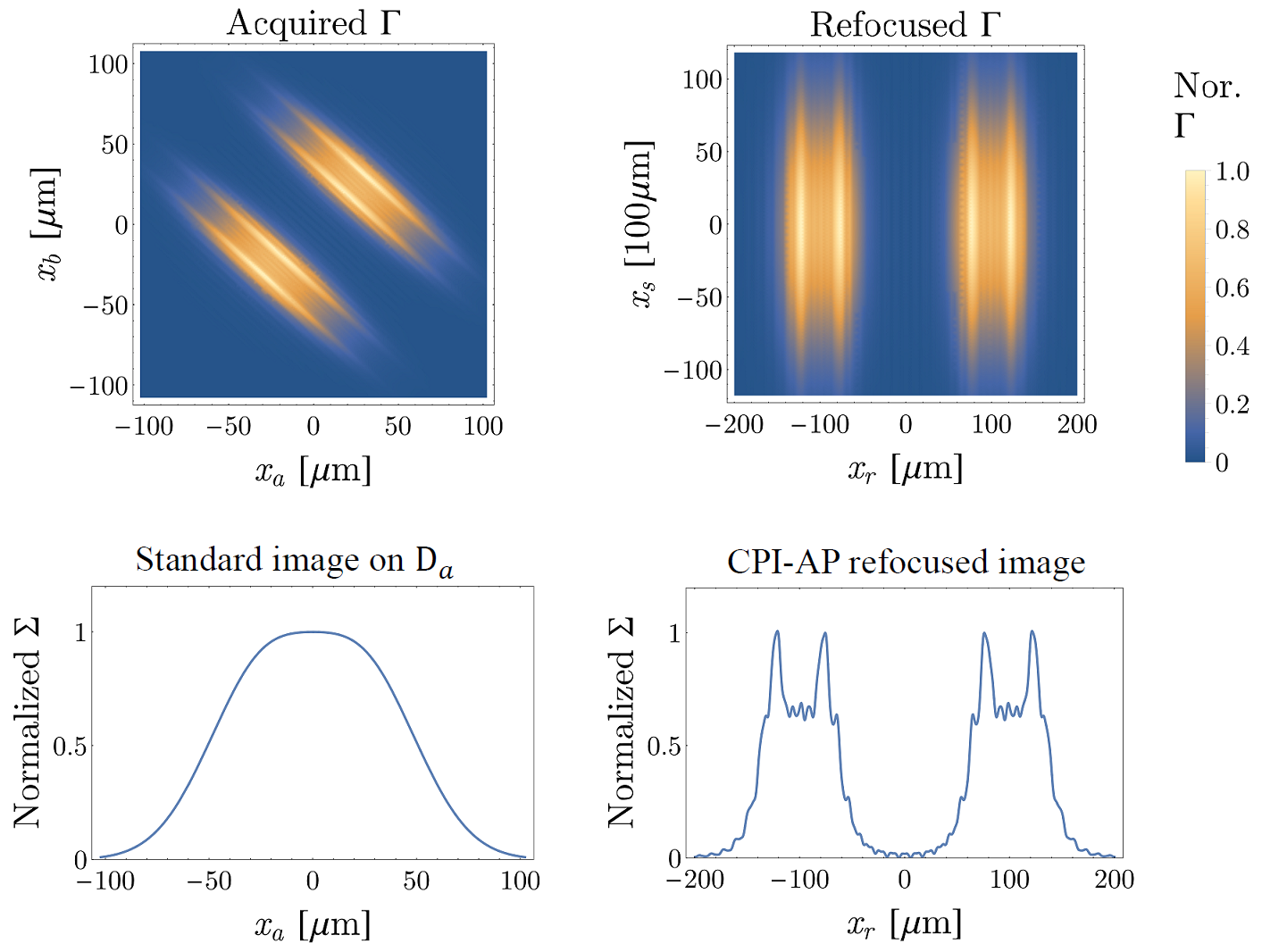}
\caption{\textit{Top left.} Simulation of the correlation function of Eq.\ref{eq:cnp:gamma} associated with the chaotic light CPI-AP setup of Figure~\ref{fig:cnp:setup}, obtained by evaluating the pixel-by-pixel correlation between $\text{D}_a$ and $\text{D}_b$, and then projecting on the $x$-axis. The object is a double-slit mask with center-to-center slit distance twice the slit width; see text for further details. The $x$-projection of the image observed on detectors $\text{D}_a$ and $\text{D}_b$, separately, can be obtained by integrating on $x_b$ and $x_a$, respectively. In the \textit{bottom left} panel, integration has been performed with respect to $\br_b$: the image retrieved by $\text{D}_a$ is clearly out of focus. \textit{Top right.} Refocused correlation function obtained from $\Gamma$, in the top left panel, through the linear transformation of variables defined in Eq.~\eqref{eq:cnp:gammaref}. \textit{Bottom right.}  Integration with respect to $\br_s$ of the refocused correlation function $\Gamma_{\text{ref}}$ in the top right panel. The image of the double-slit mask is now properly resolved.}
\label{fig:cnp:gamma}
\end{figure}

In order to develop a feeling about the result of Eq.~\eqref{eq:cnp:gamma}, we represent, in the upper left panel of Fig.~\ref{fig:cnp:gamma}, a numerical evaluation of the correlation function obtained by considering: the object as an emitter of chaotic light of wavelength $\lambda = 480\,\text{nm}$, consisting in a double-slit mask with center-to-center distance $d=200\,\mu\text{m}$ and width $d/2$, placed in $z_m=(z_a+z_b)/2=290\,\mathrm{mm}$ (i.e., at equal distances from the planes $\text{D}^o_a$ and $\text{D}^o_b$); the lens $L_f$ with focal length $f=58\,\text{mm}$ and numerical aperture $\text{NA}=0.08$, as seen from the object plane; the distance between the two imaged planes as
$|z_b-z_a|=10\,\text{mm}$. Notice that $z_a$ and $z_b$ has been chosen in such a way that the images separately retrieved by detectors $\text{D}_a$ and $\text{D}_b$ are outside the depth of field of lens $L_f$, and are thus out-of-focus.
The density plot of the correlation function reported in the top left panel of Figure~\ref{fig:cnp:gamma} shows that the information about the double-slit mask is contained along the bisector of the plane $x_a, x_b$. As we shall better see later, this is a consequence of the ``one-to-one'' correspondence between the points $(\bm\rho_a,\bm\rho_b)$ of the correlation function, and the rays captured by $L_f$ that cross both the plane $\text{D}^o_a$, in $\bm\rho_a/M_a$, and the plane $\text{D}^o_b$, in $\bm\rho_b/M_b$. 

To clarify these concepts and the imaging properties of the correlation function $\Gamma(\br_a,\br_b)$, we shall consider the geometrical-optics limit $k\to\infty$, in which the most relevant contribution to the integral in Eq.~\eqref{eq:cnp:gamma} can be evaluated by applying the method of stationary phase \cite{saleh2007fundamentals,schulman2012techniques}. Actually, the stationary points of the phase $k(\phi_a-\phi_b)$, appearing in Eq.~\eqref{eq:cnp:gamma},
with respect to $\bm\rho_{\ell}$, $\bm\rho_{\ell'}$ and $\bm\rho_o$ enable us to determine the geometrical correspondence between points on the object and points on the sensors $\text{D}_a$ and $\text{D}_b$, providing the the dominant asymptotic contribution to the correlation function:
\begin{equation}\label{eq:gamma_geom_c}
\Gamma(\bm\rho_a,\bm\rho_b) \sim
\A^2 \left[
\frac{1}{z_b-z_a}\left(
\frac{z-z_a}{M_b}\bm\rho_b - \frac{z-z_b}{M_a}\bm\rho_a
\right)\right]  \left| P\left[
\frac{1}{z_b-z_a}\left(
\frac{z_b}{M_a}\bm\rho_a - \frac{z_a}{M_b}\bm\rho_b
\right)\right]\right|^4 .
\end{equation}
This result shows that, independent of the distance $z$ of the object mask from the lens $L_f$, in the geometrical limit, the correlation of intensity fluctuations encodes an image of both the (squared) object intensity profile $\A^2$ and the lens pupil function $P$. The dependence of $\A^2$ on both detectors coordinate explains the behaviour of the correlation function observed in in the top left panel of Figure~\ref{fig:cnp:gamma}, as already discussed. 
The image of the object depends only on the coordinate of one detector, either $\br_a$ or $\br_b$, only if the object mask lies in either one of the planes $\text{D}^o_a$ or $\text{D}^o_b$, respectively. For $z=z_a$ ($z=z_b$), $\A^2$ does not depend any longer on $\bm\rho_b$ ($\bm{\rho}_a$), and the integration of the correlation function on $\bm\rho_b$ ($\bm{\rho}_a$) gives a focused image of the object:
\begin{equation}\label{eq:cnp:sigmaa}
\Sigma_a(\bm\rho_a) = \int d^2\bm\rho_b\Gamma (\bm\rho_a,\bm\rho_b) \sim \A^2 \left( \frac{\br_a}{M_a} \right) , \quad \Sigma_b(\bm\rho_b) = \int d^2\bm\rho_a\Gamma (\bm\rho_a,\bm\rho_b) \sim \A^2 \left( \frac{\br_b}{M_b} \right)
\end{equation}
with $M_a$ ($M_b$) the transverse magnification. By working in the wave optics regime, one would find that this image has the same point-spread function and depth of field as the corresponding conventional image retrived by sensor $\mathrm{D}_a$ ($\mathrm{D}_b$) alone. However, in the more general case in which the object does not lie in either one of the conjugate planes of the detectors and is outside the DOF, as reported in the upper-left panel of Fig.~\ref{fig:cnp:gamma}, the integral of the correlation function $\Gamma$ on either one of the detector coordinates gives rise to blurred images. This is shown in the bottom-left panel of the same figure, where integration of Eq.~\eqref{eq:cnp:gamma} (or, equivalently, of Eq.~\eqref{eq:gamma_geom_c}) on $x_b$ gives rise to a blurred image of the double-slit.  

In order to decouple the image of the object from the image of the lens, thus obtaining a ``refocusing'' algorithm, we shall define proper linear combinations of the detector coordinates $\br_a$ and $\br_b$, such as the two variables
\begin{equation}
\label{eq:cnp:rhors}
\bm\rho_r = \frac{1}{z_b-z_a}\left(
\frac{z-z_a}{M_b}\bm\rho_b - \frac{z-z_b}{M_a}\bm\rho_a
\right) , \quad
\bm\rho_s = \frac{1}{z_b-z_a}\left(
\frac{z_b}{M_a}\bm\rho_a - \frac{z_a}{M_b}\bm\rho_b
\right) .
\end{equation}
By inverting the transformation in Eq.~\eqref{eq:cnp:rhors}, we obtain the refocused correlation function:
\begin{equation}\label{eq:cnp:gammaref}
\Gamma_\text{ref}(\bm\rho_r,\bm\rho_s) =
\Gamma\left[
\alpha_a \bm\rho_s + \beta_a \bm\rho_r ,
\alpha_b \bm\rho_s + \beta_b \bm\rho_r
\right] \sim 
\A\left(\bm\rho_r\right)^2
\left| P\left(\bm\rho_s \right)\right|^4
\end{equation}
with
\begin{equation}
\alpha_j = M_j \frac{z-z_j}{z}, \quad  \beta_j =  M_j \frac{z_j}{z}, \qquad \text{with } j=a,b.
\end{equation}
From the last line of Eq.~\eqref{eq:cnp:rhors}, it is evident that the performed linear transformation of the argument of $\Gamma$ realigns all the displaced images corresponding to different values of $\br_b$. This is clearly visible in the upper-right panel of Fig.~\ref{fig:cnp:gamma}, where we report the refocused correlation function obtained by ``reordering'' the correlation function of the upper-left panel according to the change of variable of Eq.~\eqref{eq:cnp:gammaref}. As shown in the bottom-left panel of Fig.~\ref{fig:cnp:gamma}, no blurring occurs anymore upon integrating the refocused correlation function over the variable $\bm\rho_s$; in fact, this integral gives the final refocused image
\begin{equation}\label{eq:cnp:sigmar}
\Sigma_\text{ref}(\bm\rho_r) =
\int d^2\bm\rho_s \Gamma_\text{ref}(\bm\rho_r,\bm\rho_s) \sim  \A\left(\bm\rho_r\right)^2 .
\end{equation}
Although any combination $\br_s$ gives a focused image of the object, the integration over $\br_s$ reported  in Eq.~\eqref{eq:cnp:sigmar} enables to exploit the whole signal collected by the two detectors, hence, to considerably increase the signal-to-noise ratio of the final image.

\begin{figure}
    \centering
    \includegraphics[width=0.6\textwidth]{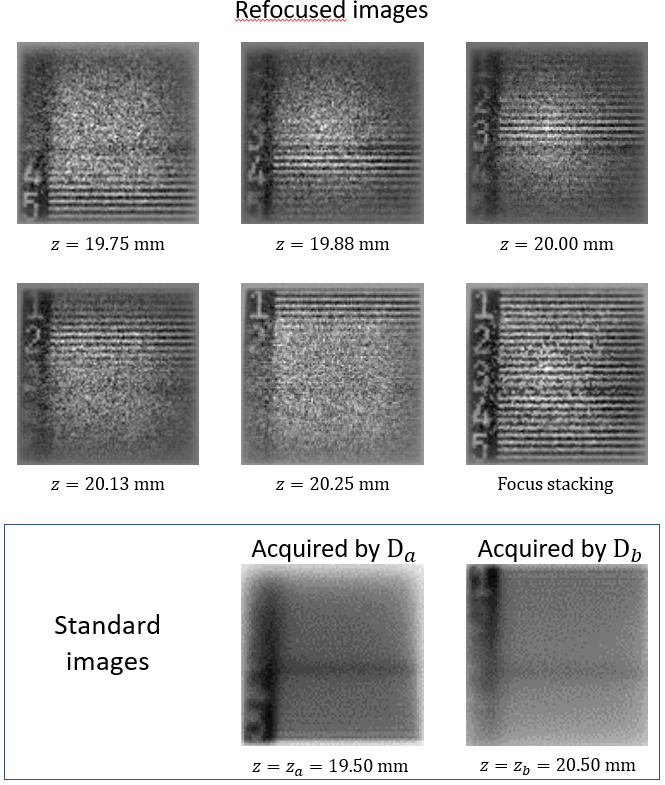}
    \caption{Refocused images of a depth-of-field target at different distances $z$ from the main lens, as enabled by chaotic light CPI-AP. The conjugate planes of the detectors are placed at a distance $z_a=19.5\,\mathrm{mm}$ and $z_b=20.5\,\mathrm{mm}$ from the lens. The two panels in the bottom row represent the standard images retrieved by $\mathrm{D}_a$ and $\mathrm{D}_b$: here, all the details of the target are blurred. Refocusing is performed by means of Eqs.~\eqref{eq:cnp:gammaref}-\eqref{eq:cnp:sigmar} at different distances $z$ between $z_a$ and $z_b$, as indicated in each panel. 
    A panel containing a stacking of all refocused images, in which all the lines and numbers are resolved, is also reported to show the DOF increase capability of the CPI-AP protocol. }
    \label{fig:simulation}
\end{figure}

A further simulation has been performed by considering a three-dimensional object such a depth of field target tilted by $3.8^{\circ}$ and characterized by equal line thickness and spacing between centers of black and white lines of $200\,\mu\mathrm{m}$. Chaotic light is simulated by randomly illuminating the object with light of transverse coherence length $\sim 100\,\mu\mathrm{m}$, and the correlation function $\Gamma$ is reconstructed, by averaging products of intensity fluctuations over $30$k frames; the results are reported in Fig.~\ref{fig:simulation}, where refocusing has been obtained through Eqs.~\eqref{eq:cnp:gammaref}-\eqref{eq:cnp:sigmar}. Stacking of all refocused images is also reported therein, thus demonstrating the DOF enhancement enabled by CPI-AP.

\section{CPI-AP with entangled photon illumination}
\label{sec:enp}

Entangled photons represent the most paradigmatic case of correlated light beams. In the mid nineties, entangled photons produced by spontaneous parametric down-conversion (SPDC) \cite{mandel1995optical} opened the way to quantum imaging \cite{pittman1996two}. Despite being less practical to produce than light from chaotic sources, there is evidence that entangled light can provide otherwise inaccessible noise reduction effects \cite{brida2009measurement,brida2010experimental}. 

\begin{figure}
\centering
\includegraphics[width=0.9\textwidth]{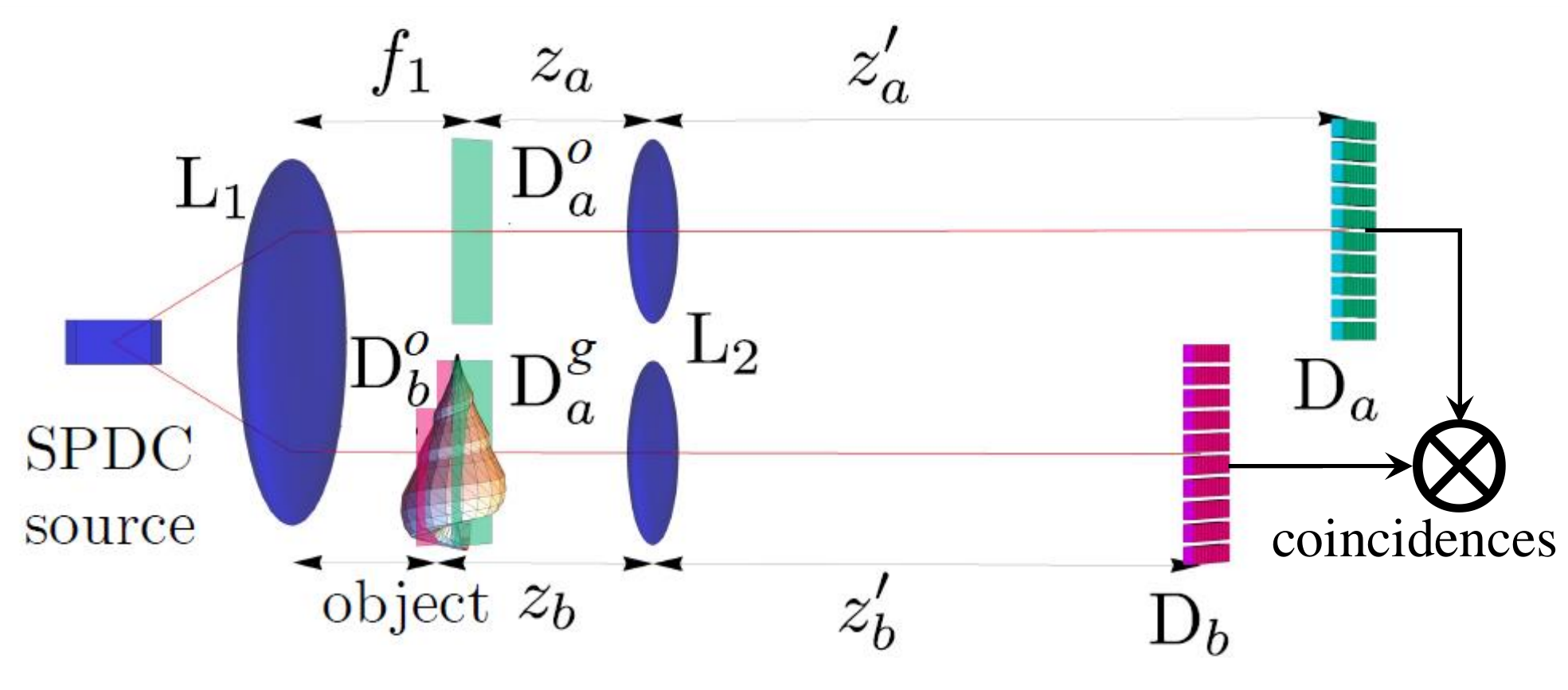}
\caption{Schematic representation of the CPI-AP protocol in the case of entangled-photon illumination. 
As opposed to the case of chaotic light illumination, depicted in Fig.~\ref{fig:cnp:setup}, here the object is not common to both light paths. The lens $L_2$ in arm $b$ focuses a standard image of the plane $\mathrm{D}^o_b$ on the detector $\mathrm{D}_b$, while an identical lens along arm $a$ is used to focus the ghost image of the plane $\mathrm{D}^g_a$ on the detector $\mathrm{D}_a$. The planes $\mathrm{D}^o_a$ and $\mathrm{D}_g^a$, conjugate to each other at second order, coincide with the focal plane of lens $L_1$.}
\label{fig:enp:setup}
\end{figure}

The proposed CPI-AP protocol adapted to entangled photons illumination is pictured in Fig.~\ref{fig:enp:setup}.  The ``signal'' ($s$) and ``idler'' ($i$) entangled photon pairs are emitted by SPDC along two different directions, and impinge on lens $L_1$, of focal length $f_1$, which collimates the incoming beams. Only one of the entangled photon beams (the one propagating along path $b$) illuminates the object of interest: we shall identify with $\mathrm{D}_a^g$ the specific plane of the object laying in the focal plane of lens $L_1$. A pair of identical lenses $L_2$, placed at a distance $f_1+z_a$ from $L_1$, enable to reproduce the \textit{ghost image} of $\mathrm{D}_a^g$ on detector $\mathrm{D}_a$, by means of correlation measurements with detector $\mathrm{D}_b$. The lenses $L_2$, of focal lengths $f_2$, serve to image on detectors $\mathrm{D}_a$ and $\mathrm{D}_b$ the planes $\mathrm{D}_a^o$ and $\mathrm{D}_b^o$, respectively, placed at a distance $z_a$ and $z_b$ from $L_2$. In fact, the distances $z'_a$ and $z'_b$ between $L_2$ and the detectors, along path $a$ and $b$, respectively, satisfy the thin lens equations: 
\begin{equation}\label{eq:enp:tl}
\frac{1}{z_i} + \frac{1}{z'_i} = \frac{1}{f_2}, \quad \text{with } i = a, b.
\end{equation}
Hence, similar to the previous case, the planes $\mathrm{D}_a^g$ and $\mathrm{D}_b^o$, are the ``arbitrary planes'' chosen within the three-dimensional scene. As we shall prove below, the coincidence counting of photon pairs detected by the two sensors $\mathrm{D}_a$ and $\mathrm{D}_b$ enable plenoptic imaging of the object of interest.

Coincidence counting is formally described by the Glauber correlation function \cite{scully1997quantum}
\begin{equation}\label{eq:G2}
G^{(2)}(\br_a,\br_b) = \bra{\Psi_2} E_a^{(-)}(\br_a)E_b^{(-)}(\br_b) E_b^{(+)}(\br_b) E_a^{(+)}(\br_a) \ket{\Psi_2}
\end{equation}
where $E_{a,b}^{(\pm)}$ are the positive- and negative-frequency contributions to the electric field component that propagate towards $\mathrm{D}_a$ and $\mathrm{D}_b$, evaluated at equal times, and
\begin{equation}\label{eq:spdc:state}
| \Psi_2 \rangle = \int d^2\bm{\kappa}_i d^2\bm{\kappa}_s h_{\mathrm{tr}} (\bm{\kappa}_i + \bm{\kappa}_s) a^{\dagger}_{\bm{k}_s} a^{\dagger}_{\bm{k}_i} |0\rangle,
\end{equation}
is the quasi-monochromatic (with wavelength $\lambda$) biphoton state. Here, the creation operators $a^{\dagger}_{\bm{k}_i} a^{\dagger}_{\bm{k}_s}$ generate a pair of photons with wavevectors $\bm{k}_s$ (signal) and $\bm{k}_i$ (idler) from the vacuum $|0\rangle$. Both wavevectors have modulus $|\bm{k}_s|=|\bm{k}_i| = k = 2\pi/\lambda$, and the variables $\bm{\kappa}_s,\bm{\kappa}_i$ represent the transverse momentum components with respect to the propagation directions of signal and idler, respectively. The function $h_{\mathrm{tr}}$ appearing in Eq.~\eqref{eq:spdc:state} is related by Fourier transform to the amplitude profile $F$ of the laser pump on the crystal:
$F(\br) = \int d^2\bm{\kappa} \, e^{i \kappa\cdot\br} h_{\mathrm{tr}}(\bm{\kappa})$. 
Notice that, though we will show the results for beams of equal central wavelength, the working principle is unchanged in case signal and idler have different wavelengths. 

Now, we compute the optical propagators of the transverse momentum components towards $\br_j$ along the path $j$, in the hypothesis that both the aperture of lens $L_1$ and the pump laser beam have infinite extension, and consider a planar object with amplitude transmission profile $A(\br_o)$ placed at an arbitrary distance $z$ before the lens $L_2$. The correlation function reads, up to irrelevant constants,
\begin{multline}\label{eq:gamma:enp}
G^{(2)}(\bm\rho_a,\bm\rho_b) =
\Biggl| \int d^2\bm\rho_o A(\bm\rho_o)
e^{-i k \frac{z_a}{z(z-2z_a)}\bm\rho_o^2} \\ \times
\int d^2\bm\rho_2 \, P_2(\bm\rho_2) \int d^2\bm\rho_2 \, P_2(\bm\rho'_2) e^{i k [ \psi_a(\bm\rho_o,\bm\rho_2,\bm\rho_a)+\psi_b(\bm\rho_o,\bm\rho'_2,\bm\rho_b) ]} \Biggr|^2,
\end{multline}
where
\begin{subequations}
\begin{align}
\psi_a (\bm\rho_o,\bm\rho_{2},\bm\rho_i) = & \,
\frac{z_a-z}{2z_a(z-2z_a)}\bm\rho_{2}^2
\label{eq:psia:enp} -
\left(\frac{\bm\rho_o}{z-2z_a}-\frac{\bm\rho_a}{M_a z_a}\right)\cdot\bm\rho_{2} , 
\\
\psi_b (\bm\rho_o,\bm\rho_{2},\bm\rho_i) = & 
\left(\frac{1}{z}-\frac{1}{z_b}\right)\frac{\bm\rho_{2}^2}{2}-
\left(\frac{\bm\rho_o}{z}-\frac{\bm\rho_b}{M_b z_b}\right)\cdot\bm\rho_{2} ,
\label{eq:psib:enp}
\end{align}
\end{subequations}
with $M_j=-z'_j/z_j$ the magnifications provided by the lenses $L_2$, on the two paths $j=a,b$. Different from the chaotic light case, where the correlation function of Eq.~\eqref{eq:cnp:gamma} depends on the object (squared) \textit{intensity profile}, the $G^{(2)}$ function associated with entangled photon CPI-AP depends on the \textit{transmission amplitude} of the object; this indicates that the retrieved plenoptic (ghost) images are coherent rather than incoherent. This difference is a consequence of the coherent nature of entangled photons, as well as of the fact that the object is not in the common path of entangled photons pairs, but is illuminated by only one of the entangled beams.
To interpret the above result, we shall notice that the term containing the quadratic phases in the lens coordinate $\br_2$ indicates two focusing conditions, one for each path. In path $b$, the focusing condition reads $z=z_b$ (i.e., $z$ satisfies the thin-lens equation \eqref{eq:enp:tl} with $i=b$), indicating that the object is directly imaged on the plane of detector $\mathrm{D}_b$. In path $a$, the focusing condition $z=z_a$ is less intuitive, since there is no object placed in this path; in fact, this condition corresponds to the situation where the ``ghost image'' of the object placed at a distance $z=z_a$ from the lens $L_2$, in arm $b$, is focused on the plane of detector  $\mathrm{D}_a$ by means of coincidence counting between $\mathrm{D}_a$ and $\mathrm{D}_b$ \cite{pittman1996two,pepe2016correlation}. Such focused ghost image is characterized by a positive magnification $-M_a=z'_a/z_a$: The minus sign is due to the double inversion of the ghost image on from $\mathrm{D}_a^g$ to $\mathrm{D}_a$. 

\begin{figure}
\centering
\includegraphics[width=0.6\textwidth]{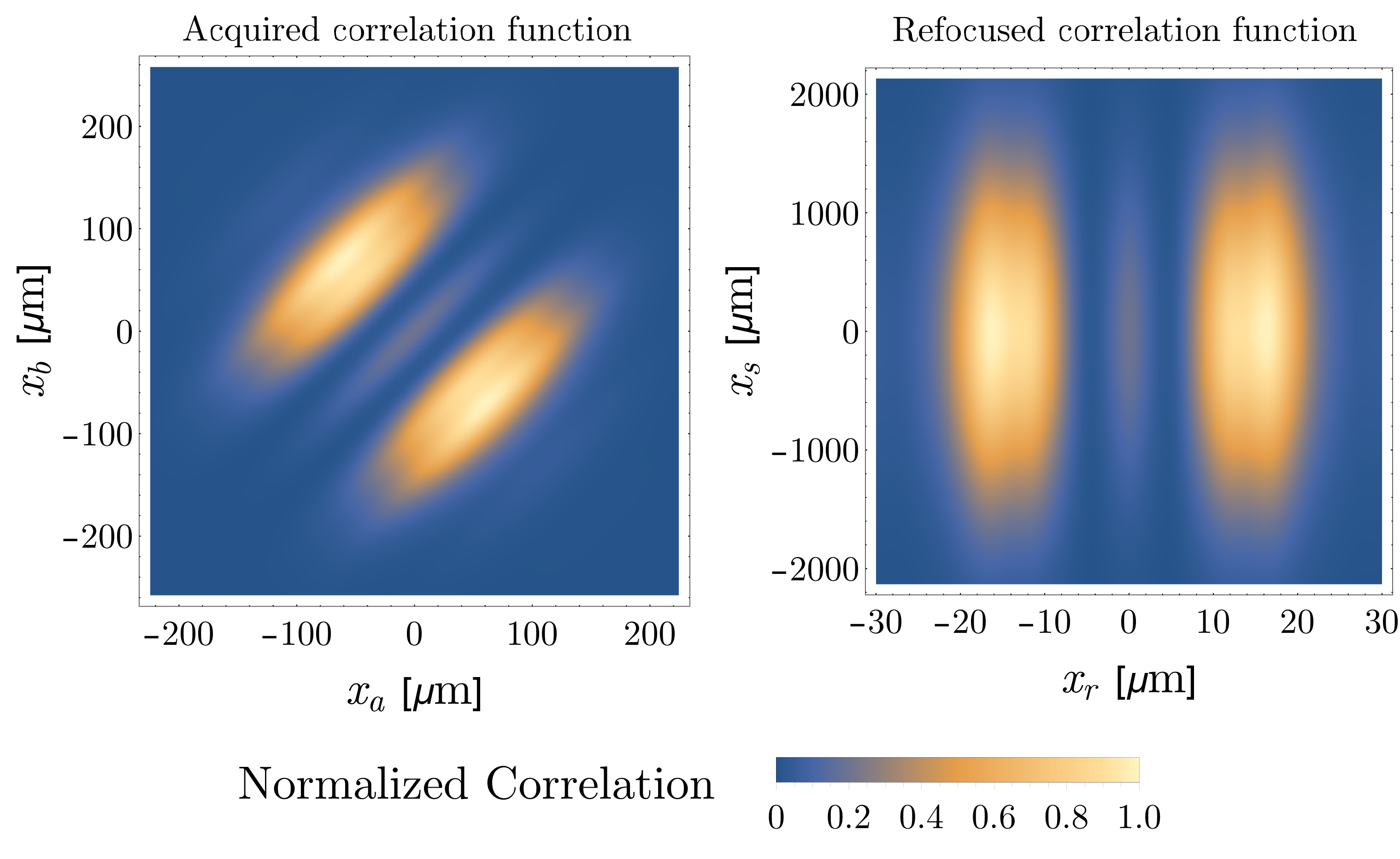}
\caption{\textit{Left panel.} Simulation of CPI-AP with entangled photon pairs, representing the projection on the horizontal coordinates of the image of a double slit, obtained by evaluating pixel-by-pixel coincidence counting between $\mathrm{D}_a$ and $\mathrm{D}_b$ as in Eq.~\eqref{eq:G2}. Details on numerical parameters are given in the text. \textit{Right panel.} Refocused correlation function, constructed according to Eq.~\eqref{eq:ref:enp}; the multiple images of the double slit, one for each value of $x_s$, are aligned parallel to the $x_s$-axis and are not blurred upon integration on this variable, as reported in Eq.~\eqref{eq:enp:sigmar}.}
\label{fig:enp:gamma}
\end{figure}

In order to show the working principle of CPI-AP in the case of entangled photons, we report in the top-left panel of Fig.~\ref{fig:enp:gamma} a numerical calculation of $G^{(2)}$ from Equation~\eqref{eq:gamma:enp}, in the case of photon wavelength $\lambda = 710\,\mathrm{nm}$, and considering an imaging lenses $L_2$ with numerical aperture $\text{NA}=0.1$ and focal length $f_2=30 \,\mathrm{mm}$. 
In analogy with the chaotic light case,  the planar sample is a double-slit mask with center-to-center distance $d=30\,\mu\mathrm{m}$ and width $a = d/2$, placed at $z_m=(z_a+z_b)/2$.
Here, the two planes $\mathrm{D}_b^g$ and $\text{D}_b^o$ are chosen to have a relative distance $|z_b-z_a|=1\,\mathrm{mm}$ larger than the DOF of the ghost image.
Similar to the results reported in Fig.~\ref{fig:cnp:gamma}, also in this case, the correlation function of CPI-AP contains information about the double slit mask, but the images are not oriented along either one of the axes. 
Due to the positive magnification $M_a$, the ghost image of the sample is not inverted, and the coherent images in Figure~ \ref{fig:enp:gamma} are inverted compared to the incoherent images in Figure~\ref{fig:cnp:gamma}. 
In close analogy with Eq.~\ref{eq:cnp:sigmaa}, if the object is placed in one of the reference plane at a distance $z=z_a$ ($z=z_b$), the integral of the correlation function over $\mathrm{D}_b$ ($\mathrm{D}_a$) yields the ghost image of the plane $\text{D}_a^g$ (the conventional image of the plane $\mathrm{D}_b^o$). In the more general case considered in Fig.~\ref{fig:enp:gamma}, both these images are heavily blurred.

By applying the stationary-phase condition to $k (\psi_a + \psi_b)$, we find that the refocused image can be obtained by re-expressing the correlation function in terms of the two variables 
\begin{equation}\label{eq:enp:rhors}
\bm\rho_r = \frac{1}{z_b-z_a}\left(
\frac{z-z_a}{M_b}\bm\rho_b + \frac{z-z_b}{M_a}\bm\rho_a
\right) , \quad
\bm\rho_s = \frac{1}{z_a-z_b}\left(
\frac{z_b}{M_a}\bm\rho_a + \frac{z_a}{M_b}\bm\rho_b
\right) ,
\end{equation}
where $\bm\rho_r$ parametrizes, as we will shortly see, a unit-magnification image of the object.
The refocused correlation function is thus obtained by inversion of Eq.~\eqref{eq:enp:rhors}, and reads
\begin{align}\label{eq:ref:enp}
G^{(2)}_\text{ref} & (\bm\rho_r,\bm\rho_s) = 
G^{(2)} (\alpha_a \bm\rho_s + \beta_a\bm\rho_r,
\alpha_b \bm\rho_s + \beta_b \bm\rho_r ) \nonumber\\ 
& \sim 
\left|A\left(\bm\rho_r\right)\right|^2 \left|P_2\left(\bm\rho_s \right)\right|^2
\left|P_2\left(\frac{2\,z_a-z}{z}\bm\rho_s - \frac{2\,z_a}{z}\bm\rho_r\right)\right|^2 ,
\end{align}
with
\begin{equation}
\alpha_a = -M_a \frac{z-z_a}{z}, \quad \beta_a = -M_a \frac{z_a}{z}, \quad
\alpha_b = M_b \frac{z-z_b}{z}, \quad \beta_b = M_b \frac{z_b}{z}
\end{equation}
In the right panel of Fig.~\ref{fig:enp:gamma} we report the result obtained by applying the refocusing algorithm of Eq.~\eqref{eq:ref:enp} to the correlation function reported in the left panel: All the images of the double-slit mask have been realligned and are now parallel to the $x_s$ axis; hence, no blurring occurs upon integration over $\bm\rho_s$. In fact, the refocused image is given by
\begin{equation}\label{eq:enp:sigmar}
\Sigma_\text{ref}(\bm\rho_r) = \int d^2\bm\rho_s G^{(2)}_\text{ref}(\bm\rho_r,\bm\rho_s) \sim \mathcal{P}(\bm\rho_r) \left|A\left(\bm\rho_r\right)\right|^2 .
\end{equation}
Unlike in CPI-AP with chaotic light, here the refocused image in the geometrical limit is modulated by an envelope
\begin{equation}
\mathcal{P}(\bm\rho_r) = \int d^2\br_s \left|P_2\left(\bm\rho_s \right)\right|^2
\left|P_2\left(\frac{2\,z_a-z}{z}\bm\rho_s - \frac{2\,z_a}{z}\bm\rho_r\right)\right|^2
\end{equation}
that depends on the aperture of the lens $L_2$. By considering, both for definiteness and for computational feasibility, a Gaussian pupil function $P_2(\br_2)=\exp(-\br_2^2/2\sigma_p^2)$ of width $\sigma_p$, the envelope function reads
\begin{equation}
\mathcal{P}(\bm\rho_r) \propto \exp\left(- \frac{\br_r^2}{\sigma_p^2 [z_a^2 + (z-z_a)^2] }  \right) ,
\end{equation}
thus reaching the smallest width (hence, the maximal disturbance to the refocused image) when $z\simeq z_a$.

\begin{figure}
\centering
\includegraphics[width=\textwidth]{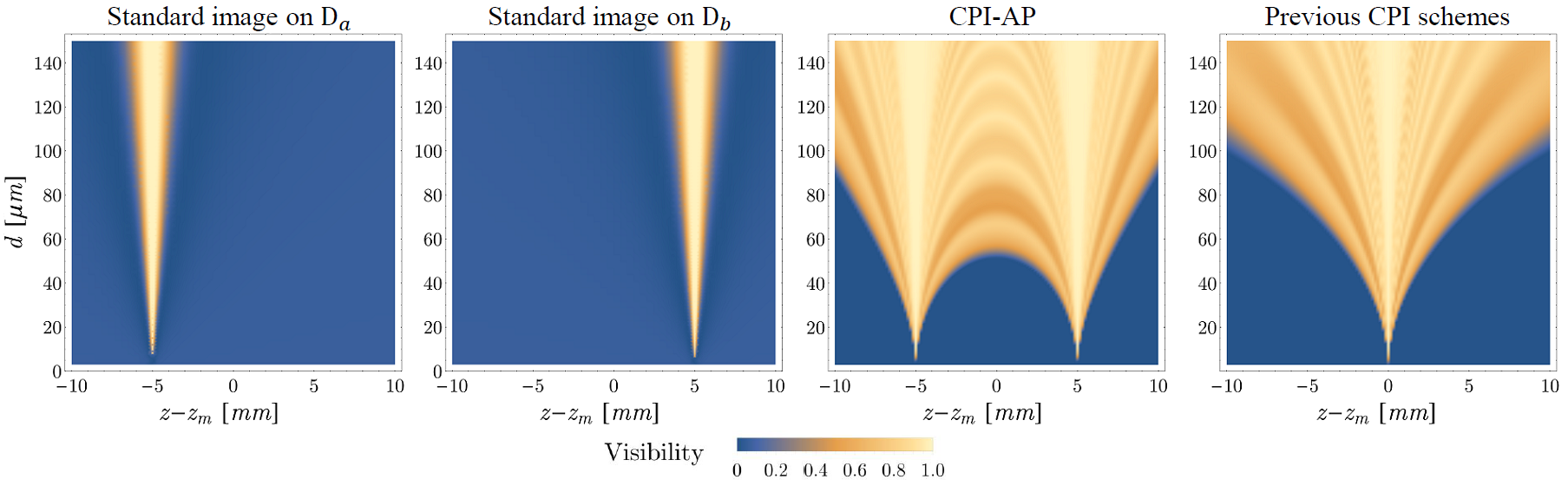}
\caption{Comparison of the resolution versus depth of field tradeoff in standard imaging (first two panels), CPI-AP, and previous CPI schemes. The comparison in made by plotting the visibilities of a double-slit mask with center-to-center distance $d$ and slit width $d/2$, obtained by the considered imaging methods, as a function of both the resolution $d$ and the axial coordinate $z$, directly related with the image DOF. The position of the mid-point $z_m=(z_a+z_b)/2$ between the two focused image planes in the CPI-AP protocol is taken as a reference. The first and second panels from the left refer to the standard images read on the detectors $\mathrm{D}_a$ and $\mathrm{D}_b$, respectively. The third panel contains the visibility of the refocused CPI-AP image. In the fourth panel, we report for comparison the visibility obtained with one of the previously developed CPI schemes, in which one of the reference planes coincides with the focusing element and the other one is placed at a distance $z_m$ from it.}
\label{fig:cnp:vis}
\end{figure}

\section{Depth-of-field enhancement}

In order to characterize the depth-of-field improvement of CPI-AP with respect to both standard imaging and previous plenoptic imaging protocols, we show in Fig.~\ref{fig:cnp:vis} the visibility of the images of double-slits masks with center-to-center distance $d$ and slit width $a=d/2$. The specific result reported in the figure is obtained by considering the setup described in Sec.~\ref{sec:cnp} and by using the same conditions employed for the simulation in Fig.~\ref{fig:cnp:gamma}. The visibility is computed according to the definition
\begin{equation}\label{eq:def:vis}
V=\frac{\Sigma(M d/2)-\Sigma(0)}{\Sigma(M d/2)+\Sigma(0)} ,
\end{equation}
with $M$ the image magnification, and comes out to depend on both the slit distance $d$, which we shall use to define the image resolution, and the longitudinal mask position $z$, giving information on the depth of field.

The first and second panels of Fig.~\ref{fig:cnp:vis} report the visibility of the images directly retrieved by $\text{D}_a$ and $\text{D}_b$, as described by Eq.~\eqref{eq:cnp:sigmaa}. The third panel shows the visibility of the CPI-AP refocused image as given by Eq.~\eqref{eq:cnp:sigmar}. It is evident that the region of high visibility extends well beyond the superposition of the high-visibility regions in the first two panels. The resolution of CPI-AP is maximal in the reference planes $z=z_a$ and $z=z_b$, where the refocused image coincides with the conventional images described by Eq.~\eqref{eq:cnp:sigmaa}. 

In Fig.~\ref{fig:cnp:vis}, the slit distance $d\simeq 52\,\mu\mathrm{m}$ is the best resolution that can be achieved by refocusing objects placed in the mid-point $z=z_m=(z_a+z_b)/2$; here, the visibility of the refocused image is $V \simeq 0.1$. The CPI-AP protocol enables refocusing objects of this size  within a range $\Delta z_{\mathrm{CPI-AP}}\simeq 14.17\,\text{mm}$, which is more than $10$ times larger than the range where the same double slit can be resolved by conventional imaging, since $\Sigma_a$ (first panel) and $\Sigma_b$ (second panel) are characterized by $\Delta z_a\simeq 1.33\,\mathrm{mm}$ and $\Delta z_b\simeq 1.38\,\mathrm{mm}$, respectively. The slight oscillations observed in the high-visibility region of the refocused image originate from the intrinsically coherent-imaging nature of CPI [see Eqs.~\eqref{eq:cnp:gamma}-\eqref{eq:gamma:enp}]. Similar results can be obtained in the case of CPI-AP with entangled photons, with the only difference that the image $\Sigma_a$ corresponds to a ghost image. 

The fourth panel of Fig.~\ref{fig:cnp:vis} enables to extend the comparison of the resolution vs DOF tradeoff of CPI-AP with the one characterizing previous CPI schemes \cite{d2016correlation,pepe2016correlation,pepe2017exploring,scagliola2020correlation}. The visibility plot in the rightmost panel is obtained by considering a CPI system with the same numerical aperture as the CPI-AP protocol, but with one reference plane chosen close to the object and the second one coinciding with the focusing element. In the case of Fig.~\ref{fig:cnp:vis}, the DOF of CPI-AP at $d\simeq18\,\mu\mathrm{m}$ is improved by approximately a factor 3 with respect to previous CPI schemes. Thus, the availability of two reference planes enables both to obtain two high-resolution images within the scene of interest and, most important, to further improve the maximum achievable DOF.

\section{Conclusions and outlook}

The improved DOF vs resolution tradeoff of CPI-AP is certainly the most striking peculiarity of this novel protocol. Another relevant advantage compared to the previously proposed CPI schemes is the fact of not requiring sharp focusing of either the light sources, as in Refs.~ \cite{d2016correlation,pepe2016correlation} or lenses, as in Refs.~\cite{pepe2017exploring,scagliola2020correlation}, a task that is not simple to implement and manage. In fact, this difference significantly simplifies both the experimental implementation and the data analysis, and does not require the use of planar sources.

Furthermore, for the chaotic-light based setup, the light propagation along two almost identical optical paths provides the possibility to exploit in the most efficient way (and without adding artificial intensity balancing or amplifications), the dynamic range of the camera, as required when the two detectors are implemented by using disjoint parts of the same sensor \cite{pepe2017correlation}.

In view of future developments, the main perspective for the configuration with chaotic light is to develop a compact CPI camera, capable of enhancing the performances of current digital cameras.  
As for the CPI-AP protocol with entangled photon illumination, the most interesting perspective is to employ it for signal-to-noise ratio optimization: the system actually shares many features with the configuration used to obtain sub-shot-noise quantum imaging \cite{brida2009measurement,brida2010experimental,samantaray2017realization}, and a preliminary and encouraging analysis of the noise reduction factor has been performed in Ref.~\cite{descisciolo2020nonclassical} by considering a setup analogous to the one presented here. The choice of the optimal measurement protocol to enable plenoptic sub-shot-noise imaging is still an open problem which we shall address in future works.

\section*{Acknowledgments}
This work was supported by QuantERA project ``Qu3D - Quantum 3D imaging at high speed and high resolution'', by Istituto Nazionale di Fisica Nucleare (INFN) project ``PICS4ME -- Plenoptic Imaging with Correlations for Microscopy and 3D Imaging Enhancement'', and by PON ARS $01\_00141$ ``CLOSE -- Close to Earth'' of Ministero dell'Istruzione, dell'Universit\`a e della ricerca (MIUR). Project Qu3D is supported by the Italian Istituto Nazionale di Fisica Nucleare from Italy, the Swiss National Science Foundation, the Greek General Secretariat for Research and Technology, the Czech Ministry of Education, Youth and Sports, under the QuantERA programme, which has received funding from the European Union's Horizon 2020 research and innovation programme.

\section*{Disclosures}
The authors declare no conflicts of interest.

%\bibliography{full-biblio}

\begin{thebibliography}{10}
\newcommand{\enquote}[1]{``#1''}

\bibitem{adelson1992single}
E.~H. Adelson and J.~Y. Wang, \enquote{Single lens stereo with a plenoptic
  camera,} {\protect\JournalTitle{IEEE transactions on pattern analysis and
  machine intelligence}} \textbf{14}, 99--106 (1992).

\bibitem{ng2005light}
R.~Ng, M.~Levoy, M.~Br{\'e}dif, G.~Duval, M.~Horowitz, and P.~Hanrahan,
  \enquote{Light field photography with a hand-held plenoptic camera,}
  {\protect\JournalTitle{Computer Science Technical Report CSTR}} \textbf{2},
  1--11 (2005).

\bibitem{broxton2013wave}
M.~Broxton, L.~Grosenick, S.~Yang, N.~Cohen, A.~Andalman, K.~Deisseroth, and
  M.~Levoy, \enquote{Wave optics theory and 3-d deconvolution for the light
  field microscope,} {\protect\JournalTitle{Optics Express}} \textbf{21},
  25418--25439 (2013).

\bibitem{xiao2013advances}
X.~Xiao, B.~Javidi, M.~Martinez-Corral, and A.~Stern, \enquote{Advances in
  three-dimensional integral imaging: sensing, display, and applications,}
  {\protect\JournalTitle{Applied optics}} \textbf{52}, 546--560 (2013).

\bibitem{prevedel2014simultaneous}
R.~Prevedel, Y.-G. Yoon, M.~Hoffmann, N.~Pak, G.~Wetzstein, S.~Kato,
  T.~Schr{\"o}del, R.~Raskar, M.~Zimmer, E.~S. Boyden \emph{et~al.},
  \enquote{Simultaneous whole-animal 3d imaging of neuronal activity using
  light-field microscopy,} {\protect\JournalTitle{Nature Methods}} \textbf{11},
  727--730 (2014).

\bibitem{ren2017fast}
M.~Ren, R.~Liu, H.~Hong, J.~Ren, and G.~Xiao, \enquote{Fast object detection in
  light field imaging by integrating deep learning with defocusing,}
  {\protect\JournalTitle{Applied Sciences}} \textbf{7}, 1309 (2017).

\bibitem{dansereau2013decoding}
D.~G. Dansereau, O.~Pizarro, and S.~B. Williams, \enquote{Decoding, calibration
  and rectification for lenselet-based plenoptic cameras,} in \emph{Proceedings
  of the IEEE conference on computer vision and pattern recognition,}  (2013),
  pp. 1027--1034.

\bibitem{adhikarla2015exploring}
V.~K. Adhikarla, J.~Sodnik, P.~Szolgay, and G.~Jakus, \enquote{Exploring direct
  3d interaction for full horizontal parallax light field displays using leap
  motion controller,} {\protect\JournalTitle{Sensors}} \textbf{15}, 8642--8663
  (2015).

\bibitem{wanner2012globally}
S.~Wanner and B.~Goldluecke, \enquote{Globally consistent depth labeling of 4d
  light fields,} in \emph{Computer Vision and Pattern Recognition (CVPR), 2012
  IEEE Conference on,}  (IEEE, 2012), pp. 41--48.

\bibitem{ng2005fourier}
R.~Ng, \enquote{Fourier slice photography,} {\protect\JournalTitle{ACM
  Transactions on Graphics}} \textbf{24}, 735--744 (2005).

\bibitem{georgiev2012multifocus}
T.~Georgiev and A.~Lumsdaine, \enquote{The multifocus plenoptic camera,} in
  \emph{Digital Photography VIII,}  vol. 8299 (International Society for Optics
  and Photonics, 2012), p. 829908.

\bibitem{goldlucke2015plenoptic}
B.~Goldl{\"u}cke, O.~Klehm, S.~Wanner, and E.~Eisemann, \enquote{Plenoptic
  cameras,} {\protect\JournalTitle{Digital Representations of the Real World:
  How to Capture, Model, and Render Visual Reality, eds. M. Magnor, O. Grau, O.
  Sorkine-Hornung, and C. Theobalt (CRC Press, 2015)}}  (2015).

\bibitem{waller2012phase}
L.~Waller, G.~Situ, and J.~W. Fleischer, \enquote{Phase-space measurement and
  coherence synthesis of optical beams,} {\protect\JournalTitle{Nature
  Photonics}} \textbf{6}, 474 (2012).

\bibitem{georgiev2006spatio}
T.~Georgiev, K.~C. Zheng, B.~Curless, D.~Salesin, S.~K. Nayar, and C.~Intwala,
  \enquote{Spatio-angular resolution tradeoffs in integral photography.}
  {\protect\JournalTitle{Rendering Techniques}} \textbf{2006}, 263--272 (2006).

\bibitem{perez2014super}
J.~P{\'e}rez, E.~Magdaleno, F.~P{\'e}rez, M.~Rodr{\'\i}guez, D.~Hern{\'a}ndez,
  and J.~Corrales, \enquote{Super-resolution in plenoptic cameras using fpgas,}
  {\protect\JournalTitle{Sensors}} \textbf{14}, 8669--8685 (2014).

\bibitem{li2016scalable}
Y.~Li, M.~Sj{\"o}str{\"o}m, R.~Olsson, and U.~Jennehag, \enquote{Scalable
  coding of plenoptic images by using a sparse set and disparities,}
  {\protect\JournalTitle{IEEE Transactions on Image Processing}} \textbf{25},
  80--91 (2016).

\bibitem{d2016correlation}
M.~D'Angelo, F.~V. Pepe, A.~Garuccio, and G.~Scarcelli, \enquote{Correlation
  plenoptic imaging,} {\protect\JournalTitle{Physical Review Letters}}
  \textbf{116}, 223602 (2016).

\bibitem{pepe2016correlation}
F.~V. Pepe, F.~Di~Lena, A.~Garuccio, G.~Scarcelli, and M.~D'Angelo,
  \enquote{Correlation plenoptic imaging with entangled photons,}
  {\protect\JournalTitle{Technologies}} \textbf{4}, 17 (2016).

\bibitem{pepe2017exploring}
F.~V. Pepe, O.~Vaccarelli, A.~Garuccio, G.~Scarcelli, and M.~D'Angelo,
  \enquote{Exploring plenoptic properties of correlation imaging with chaotic
  light,} {\protect\JournalTitle{Journal of Optics}} \textbf{19}, 114001
  (2017).

\bibitem{scagliola2020correlation}
A.~Scagliola, F.~Di~Lena, A.~Garuccio, M.~D'Angelo, and F.~V. Pepe,
  \enquote{Correlation plenoptic imaging for microscopy applications,}
  {\protect\JournalTitle{{P}hysics {L}etters {A}}} \textbf{384}, 126472 (2020).

\bibitem{dilena2018correlation}
F.~Di~Lena, F.~Pepe, A.~Garuccio, and M.~D'Angelo, \enquote{Correlation
  plenoptic imaging: An overview,} {\protect\JournalTitle{Applied Sciences}}
  \textbf{8}, 1958 (2018).

\bibitem{pepe2017diffraction}
F.~V. Pepe, F.~Di~Lena, A.~Mazzilli, E.~Edrei, A.~Garuccio, G.~Scarcelli, and
  M.~D'Angelo, \enquote{Diffraction-limited plenoptic imaging with correlated
  light,} {\protect\JournalTitle{Physical Review Letters}} \textbf{119}, 243602
  (2017).

\bibitem{scala2019signal}
G.~Scala, M.~D'Angelo, A.~Garuccio, S.~Pascazio, and F.~V. Pepe,
  \enquote{Signal-to-noise properties of correlation plenoptic imaging with
  chaotic light,} {\protect\JournalTitle{{P}hysical {R}eview {A}}} \textbf{99},
  053808 (2019).

\bibitem{scala2020signal}
G.~Scala, G.~Massaro, M.~D'Angelo, A.~Garuccio, S.~Pascazio, and F.~V. Pepe,
  \enquote{Signal-to-noise ratio in {C}orrelation {P}lenoptic {I}maging,}
  {\protect\JournalTitle{{P}roceedings of {SPIE}}} \textbf{11347}, 1134713
  (2020).

\bibitem{georgiev2009high}
T.~G. Georgiev, A.~Lumsdaine, and S.~Goma, \enquote{High dynamic range image
  capture with plenoptic 2.0 camera,} in \emph{Frontiers in Optics 2009/Laser
  Science XXV/Fall 2009 OSA Optics \& Photonics Technical Digest,}  (Optical
  Society of America, 2009), p. SWA7P.

\bibitem{mandel1995optical}
L.~Mandel and E.~Wolf, \emph{Optical coherence and quantum optics} (Cambridge
  university press, 1995).

\bibitem{goodman2005introduction}
J.~W. Goodman, \emph{Introduction to Fourier optics} (Roberts and Company
  Publishers, 2005).

\bibitem{saleh2007fundamentals}
B.~E. Saleh and M.~C. Teich, \emph{Fundamentals of photonics} (Wiley Series in
  Pure and Applied Optics, Wiley, 2007).

\bibitem{schulman2012techniques}
L.~S. Schulman, \emph{Techniques and applications of path integration} (Courier
  Corporation, 2012).

\bibitem{pittman1996two}
T.~Pittman, D.~Strekalov, D.~Klyshko, M.~Rubin, A.~Sergienko, and Y.~Shih,
  \enquote{Two-photon geometric optics,} {\protect\JournalTitle{Physical Review
  A}} \textbf{53}, 2804 (1996).

\bibitem{brida2009measurement}
G.~Brida, L.~Caspani, A.~Gatti, M.~Genovese, A.~Meda, and I.~R. Berchera,
  \enquote{Measurement of sub-shot-noise spatial correlations without
  background subtraction,} {\protect\JournalTitle{Physical Review Letters}}
  \textbf{102}, 213602 (2009).

\bibitem{brida2010experimental}
G.~Brida, M.~Genovese, and I.~R. Berchera, \enquote{{Experimental realization
  of sub-shot-noise quantum imaging},} {\protect\JournalTitle{Nature
  Photonics}} \textbf{4}, 227--230 (2010).

\bibitem{scully1997quantum}
M.~O. Scully and M.~S. Zubairy, \emph{Quantum optics} (Cambridge university
  press, 1997).

\bibitem{pepe2017correlation}
F.~V. Pepe, F.~Di~Lena, A.~Mazzilli, A.~Garuccio, G.~Scarcelli, and
  M.~D'Angelo, \enquote{Correlation plenoptic imaging,} in \emph{2017
  Conference on Lasers and Electro-Optics Europe European Quantum Electronics
  Conference (CLEO/Europe-EQEC),}  (IEEE, 2017), pp. 1--1.

\bibitem{samantaray2017realization}
N.~Samantaray, I.~Ruo-Berchera, A.~Meda, and M.~Genovese, \enquote{Realization
  of the first sub-shot-noise wide field microscope,}
  {\protect\JournalTitle{Light: Science \& Applications}} \textbf{6}, e17005
  (2017).

\bibitem{descisciolo2020nonclassical}
E.~De~Scisciolo, F.~Di~Lena, A.~Scagliola, A.~Garuccio, F.~V. Pepe, A.~Avella,
  I.~Ruo-Berchera, and M.~D'Angelo, \enquote{Nonclassical noise features in a
  correlation plenoptic imaging setup,} {\protect\JournalTitle{{I}nternational
  {J}ournal of {Q}uantum {I}nformation}} \textbf{17}, 1941017 (2020).

\end{thebibliography}

\end{document}